\documentstyle{l-aa}

\def\Msol{\,M_\odot}

\def\kms{kms$^{-1}$}

\def\teff{$T_{\rm eff}$~}
\def\C12C13{$^{12}$C/$^{13}$C}

\def\sixt`enth{\textstyle{1\over{16}}}
\begin{document}
\input{psfig.tex}


\thesaurus{ 7(08.01.1 - 02.14.1 - 10.05.1) }
\title{ Abundances of light elements in metal-poor stars. IV. }
\subtitle{ [Fe/O] and [Fe/Mg] ratios and the history of star formation
in the solar neighborhood }
\author{ R.~G. Gratton$^1$, E. Carretta$^{1}$, F. Matteucci$^2$, 
C. Sneden$^3$ } 
\offprints{ R.G. Gratton }
\institute{ 
$^1$Osservatorio Astronomico di Padova, Vicolo dell'Osservatorio, 5,
    I-35122 Padova, ITALY\\
$^2$SISSA, Trieste, ITALY\\
$^3$University of Texas at Austin and McDonald Observatory, U.S.A. }

\date{}

\maketitle
\markboth{R.~Gratton et al.} {Abundances of light elements in metal-poor stars.
IV} 

\begin{abstract}

The accurate O, Mg and Fe abundances derived in previous papers of this series
from a homogenous reanalysis of high quality data for a large sample of stars
are combined with stellar kinematics in order to discuss the history of star
formation in the solar neighborhood. We found that the Fe/O and Fe/Mg abundance
ratios are roughly constant in the (inner) halo and the thick disk; this means
that the timescale of halo collapse was shorter than or of the same order of
typical lifetime of progenitors of type Ia SNe ($\sim 1$~Gyr), this conclusion
being somewhat relaxed (referring to star formation in the individual 
fragments) in an accretion model for the Galaxy formation. Both Fe/O and Fe/Mg
ratios raised by $\sim 0.2$~dex while the O/H and Mg/H ratios hold constant
during the transition from the thick to thin disk phases, indicating a sudden
decrease in star formation in the solar neighbourhood at that epoch. These
results are discussed in the framework of current views of Galaxy formation;
they fit in a scenario where both dissipational collapse and accretions were
active on a quite similar timescale. 

\end{abstract}

\section {Introduction}

Observations of our and outer galaxies allowed to identify various galactic
populations: the halo, the thick disk, the thin disk, and the bulge. A model
for the evolution of galaxies should explain the origin and properties of these
populations, as well as other basic observations like e.g. the relation of
Hubble types with local environment, in a unifying scheme. Current models for
galaxy formation broadly divide into two families: those considering a
dissipational collapse (Eggen et al. 1962; Larson 1974); and those which
consider galaxies as the results of the accretion of individual fragments
undergoing (some) indipendent chemical and dynamical evolution (Toomre \&
Toomre 1972; Searle \& Zinn 1978). The transition between the halo and disk
phases is continuous in smooth {\it dissipational collapse} models, while disk
formation is a secondary mechanism in {\it accretion} ones. Separation between
these two classes of models may be quite artificial: in fact various
properties of galaxies, like e.g. the light distribution of ellipticals, are
well reproduced by inhomogenous collapses leading to some kind of violent
relaxation (Lynden-Bell 1967); on the other side, simulations based on
cosmologies dominated by cold dark matter predict that in high density regions
galaxies form hierchically by merging of smaller subunits, while in low density
ones they form more gradually by infall of diffuse matter (Frenk et al. 1985).
Within this framework, the mechanisms of formation of our own galaxy (the Milky
Way) could be determined by examining fossil remnants of the early phases
represented by the old (and often metal-poor) stars. The interpretation of the
large amount of data gathered in the last years on dynamics and
metallicities (as defined by the most easily observed element, Fe) of field
stars is however still controversial, and while e.g. some authors consider the
thick disk and the bulge (Gilmore et al. 1989) as distinct galactic components,
others (Norris 1993) think they are simply the outer (and oldest) part of the
disk and central part of the halo respectively. Scenarios of galactic evolution
including a hiatus between the formation of the halo and of a secondary disk
(Ostriker \& Thuan 1975), that were introduced to justify the rarity of
metal-poor stars in the solar neighbourhood (Schmidt 1963), are widely applied
e.g. to explain the hot, metal-rich intergalactic gas seen in clusters (Berman
\& Suchkov 1991); however, up to now the observational basis for this hiatus
(based on the age gap between open and globular clusters: Demarque et al. 1992,
Carraro et al. 1999; the white dwarf cooling sequence: Wonget et al. 1987,
Knox et al. 1999; and the Th/Nd ratio nucleo-chronometer for disk and halo 
stars: Malaney \& Fowler 1989, Cowan et al. 1999) are rather weak and
controversial.

Relative abundances of O and Fe in stars of different overall metal abundance
provide further constraints to the early evolution of the halo and the
formation of the galactic disk (Wheeler et al. 1989). O is the main product of
hydrostatic He-burning: hence the ejecta of core-collapse supernovae (SNe)
resulting from the evolution of massive stars, usually identified with type~II
SNe, are expected to be very rich in O (Woosley \& Weaver 1986; Thielemann et
al 1990). On the other side, while a fraction of the Fe presently observed in
the interstellar medium was synthesized in massive stars (Thielemann et al
1990), a large fraction of it was likely produced in explosive burning under
degenerate conditions in type~Ia SNe (Nomoto et al. 1984). Typical lifetimes of
the progenitors of type~Ia SNe ($\sim 10^8\div 10^9$~yr) are much longer than
those of the progenitors of type~II SNe ($\sim 10^7$~yr), and they are actually
longer than, or of the same order of, the free fall time in the Galaxy ($\sim
3\,10^8$~yr); for these reasons the production of the bulk of Fe is expected to
be delayed with respect to that of O (Matteucci \& Greggio 1986). A clear break
in the run of O abundances with overall metallicity [Fe/H]\footnote{In this
paper we adopt the standard spectroscopic notation: [X]=log$_{10}$(X)$_{\rm
star}-$log$_{10}$(X)$_{\rm Sun}$\ for any abundance ratio X.} should signal the
onset of the contribution by type Ia SNe, and the location of this break
provides an independent estimate for the timescale of star formation during the
early stages of galactic evolution (Matteucci \& Fran\c cois 1992: hereinafter
MF). It should be added that other $\alpha-$elements (like Mg, Si, and Ca) are
expected to behave similarly to O, although for Si and Ca a small contribution
by type~Ia SNe is also expected. 

In the last years various investigations have been devoted to the study of the
run of [O/Fe] with [Fe/H] in halo and disk stars (Wheeler et al. 1989, King
1994, Nissen \& Schuster 1997, Fuhrman 1998, 1999, Israelian et al. 1998,
Boesgaard et al. 1999). However, a variety of basic questions still lacks of a
clearcut answer. The [O/Fe] ratio in the halo and the location of the change
of slope in the run [O/Fe] vs [Fe/H] have been addressed by King (1994), who
concluded that this change may occur at any value in the range
$-1.7<$[Fe/H]$<-1.0$, corresponding to timescales for the halo formation
between $3\,10^8$\ and $3\,10^9$~yr (MF); this range is large enough to
accomodate both a fast, ordered dissipational collapse (Eggen et al. 1962), or
a much slower, accretion scenario (Searle \& Zinn 1978). Edvardsson et al.
(1993) studied the [O/Fe] run in disk stars; they suggested that this ratio is
constant for [Fe/H]$>-0.2$, and argued that the spread in [Fe/H] values at any
age is an evidence for infall of metal-poor material. Even less understood is
the [O/Fe] run at intermediate metallicities, corresponding to the thick disk
phase (Gilmore et al. 1989; Nissen \& Schuster 1997).

The main concerns in previous investigations on O abundances relate (i) to the
paucity of samples of significant sizes studied in a homogeneous way, and then
to the possible existence of systematic offsets between different sets of data;
and (ii) to the discrepancy between O abundances determined using high
excitation permitted and low excitation forbidden lines (the first usually
observed in dwarfs, the second in giants). Most of this discrepancy can be
removed by adopting higher temperatures in the analysis of dwarfs (King 1993);
furthermore, the effects of departures from the Local Thermodynamic
Equilibrium (LTE) assumption when considering
the formation of high excitation permitted O~I lines should also be considered,
in order to provide abundances at the level of accuracy required for the
present purposes. Recent results based on the OH band at the extreme UV edge
of ground-based observations have further complicated this issue, suggesting
the presence of a quite strong slope in the [Fe/O] run with [Fe/H] amongst
metal-poor stars (Israelian et al. 1998, Boesgaard et al. 1999)

Both a high temperature scale, and consideration of departures from LTE were
included in the new homogeneous determinations of abundances of light elements
and Fe for a large sample of stars we are presenting in this series of papers.
A new, hopefully improved temperature scale based on IRFM temperatures for
population I stars and the new model atmospheres by Kurucz (1993, CD-ROM 13)
was obtained in Gratton et al. (1996a, Paper I). An extensive discussion of
the effects of departures from the assumption of LTE in line formation in the
stellar atmospheres was given in Gratton et al. (1999, Paper II); in that
discussion, we exploited an empirical calibration of the poorly known cross
sections for collisions with H~I atoms drawn from a parallel analysis of the
spectra of RR~Lyare variables at minimum light, where non-LTE effects are much
larger than in the stars here considered (Clementini et al. 1995). Our final
abundances for about 300 stars were presented and discussed in Carretta et al.
(2000, Paper III). We found that most discrepancies present in earlier works
have been removed, our results showing a high degree of internal consistency,
at least for stars with effective temperature \teff$>4600$~K, although our
results are still not easy to be reconciled with the O abundances from the UV
OH bands. However, in this paper we will show that once combined with stellar
kinematics and compared with models of galactic chemical evolution, our
results allow to throw new light into some of above mentioned questions: we
find that in the framework of homogeneous models, the collapse of the halo and
the formation of the thick disk occurred on a short timescale (a few
$10^8$~yr), although star formation in these environments likely lasted for
$1\div 2$~Gyr; and that there was a sudden decrease in star formation between
the thick and thin disk phases, which are then clearly distinct galactic
components. It is worth noticing that a decrease in the star formation rate
between the formation of the spheroidal components and the disc in disc
dominated galaxies, had already been suggested by Larson (Larson et al. 1976).
He suggested that such a decrease in the star formation could be due either to
the action of tidal forces inhibiting star formation during the later stages
of the collapse or to a two-phase structure of the gas, with dense clouds
forming rapidly in a spheroidal component and less dense intercloud gas not
forming stars and settling to a disc. We argue that our results fit in a
scenario in which both collapse and accretion were important in the formation
of the Milky Way, these mechanisms having similar timescales; the relative
weights of the two contributions in other galaxies might explain the Hubble
sequence.

An early, short presentation of the content of this paper was given as a talk
at the conference on Formation of the Galactic Halo (Gratton et al. 1996b).
Here we give a more complete presentation of our arguments. In the meanwhile,
Fuhrmann (1998) reached quite similar conclusions, based on an independent
careful analysis of a smaller smaple of nearby stars.

\begin{figure}
\psfig{figure=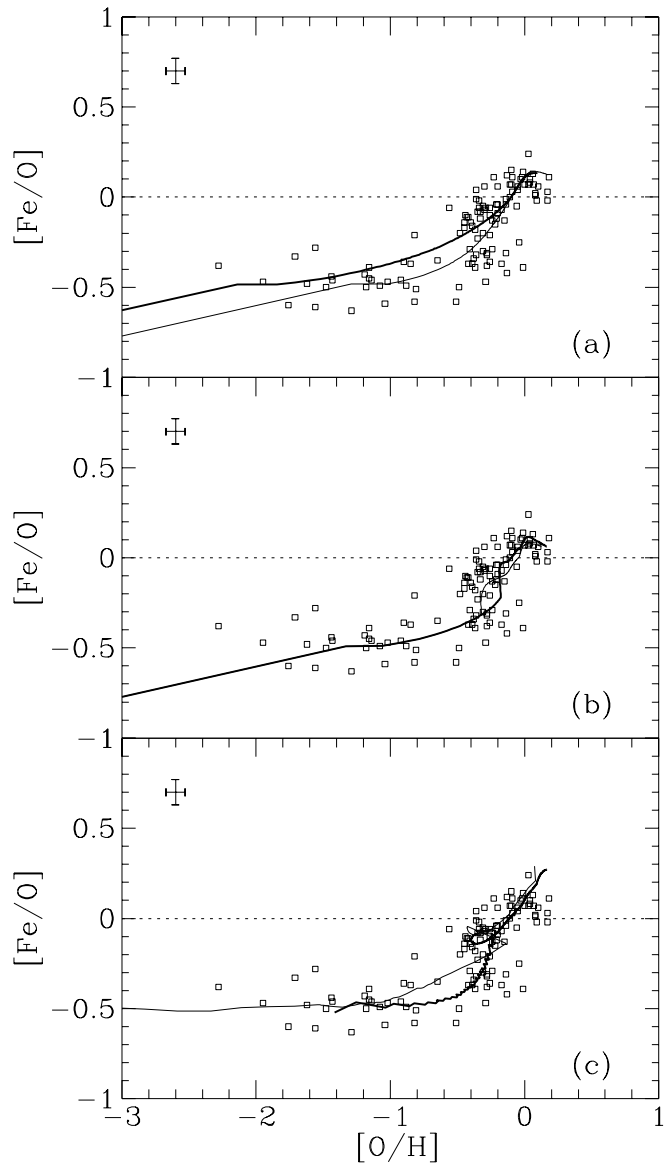,width=8.8cm,clip=}
\caption[ ]{ (a) Run of [Fe/O] ratio with [O/H] for field stars with \teff
$>4600$~K. Overimposed lines are the predictions from the models of
chemical evolution of Matteucci \& Fran\c cois 1992): model 1: thick line;
model 2: thin line. Error bar is at top left. (b) The same as panel (a) but
with models computed assuming that infall is the sum of two exponentials
("halo" and "disk"), both starting at $t=0$\ but with different decay time; in
these models, the star formation rate (SFR) has been suddenly decreased after
1~Gyr (thin line) and 2~Gyr (thick line). (c) The same as panel (b), but with
the "disk" infall starting after 2~Gyr and with a threshold of
7~$\Msol$~pc$^{-2}$\ for the SFR; thin line represents predictions of a model
where the same SFR has been assumed for both halo and disk phases, while the
thick line represents a model where the halo SFR has been encreased by an order
of magnitude} 
\label{fig:1}
\end{figure}

\begin{figure}
\psfig{figure=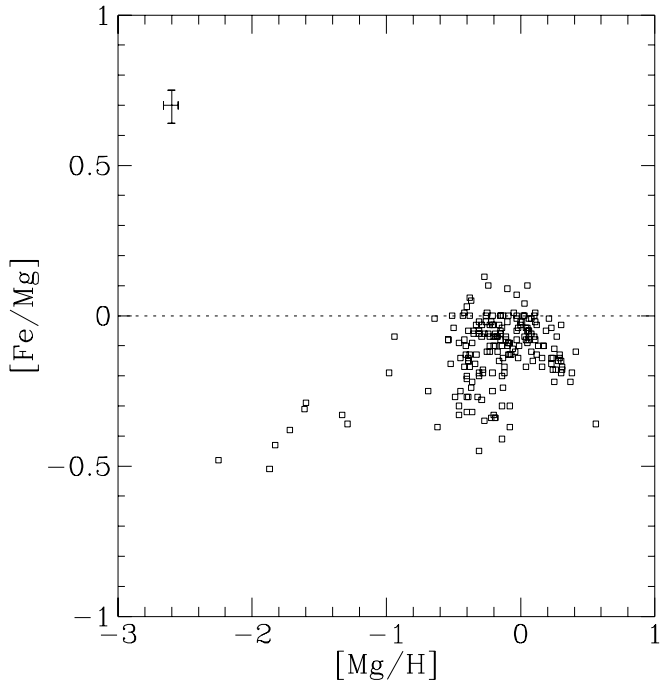,width=8.8cm,clip=}
\caption[ ]{Run of [Fe/Mg] ratio with [Mg/H] for field stars with \teff
$>4600$~K. Error bar is at top left} 
\label{fig:2}
\end{figure}

\section{ Results }

Our results for Fe, O and Mg are displayed in Fig.~\ref{fig:1} and \ref{fig:2}.
To further improve homogeneity, we only plotted data for dwarfs (Edvardsson et
al. 1993; Tomkin et al. 1992; Nissen \& Edvardsson 1992; Zhao \& Magain 1990);
all these stars have \teff$>4600$~K. Inclusion of cooler and/or lower gravity
stars in our sample does not change the present discussion. The original data
(equivalent widths) used in these papers are very consistent with each other,
and were measured on high signal-to-noise, high resolution spectra; typical
errors of individual equivalent widths determined from independent estimates
for the same stars, are $\pm 2$~m\AA, yielding errors of $\pm 0.04$~dex in
abundances derived from individual lines. Possible errors in the atmospheric
parameters required in the analysis of individual stars ($\pm 80$~K in the
effective temperatures \teff; $\pm 0.3$~dex in the surface gravity; $\pm
0.3$~\kms\ in the microturbulent velocity) cause uncertainties of $\pm
0.08$~dex in [Fe/H], $\pm 0.07$~dex in [O/H], and $\pm 0.07$~dex in [Fe/O];
these are internal errors. Analogous values for Mg are $\pm 0.05$~dex in
[Mg/H], and $\pm 0.06$~dex in [Fe/Mg]. Systematic errors should be small since
the present analysis is differential with respect to the Sun. In total, we
determined [O/Fe] ratios for about 160 stars and [Mg/Fe] for 197 stars. 

Following Wheeler et al. (1989) we have chosen to use O and Mg as reference
elements, because they are almost uniquely produced in massive stars. The
scatter of data for individual stars is small and compatible with observational
errors for stars with [O/H]$>-0.5$\ and [Mg/H]$>-0.6$, although we cannot 
exclude a gentle ($\sim 0.1$) slope in the run of [Fe/O] with [O/H]; the
scatter we obtain for more metal-poor stars (0.10~dex in both [Fe/O] and
[Fe/Mg] ratios; for O the peculiar N-rich dwarf HD~74000 was omitted) may
indicate that errors are larger in this abundance range (perhaps due to the
apparent star faintness). However, the composition of the interstellar matter
could have been slightly inhomogeneous during the early phases of galactic
evolution, as suggested by the large spread in the abundances of n-capture
elements found by McWilliam et al. (1995) for [Fe/H]$<-2$ (i.e. [O/H$<-1.5$),
and the scatter in the Fe/O ratios amongst halo stars found in the careful
analysis by Nissen \& Schuster (1997). The upper limit for intrinsic
star-to-star variations derived from the spread in our data is 0.07~dex, once
observational errors are taken into account.

Fig.~\ref{fig:1}a indicates that the run of [Fe/O] with [O/H] is quite flat
from [O/H]=$-$2.2, [Fe/O]=$-$0.5 (these are the most metal-poor stars in our
sample) to [O/H]=$-$0.29, [Fe/O]=$-$0.36, where it is possible to locate (with
an uncertainty of about $\pm 0.1$~dex) the change of slope due to the onset of
the contribution to nucleosynthesis by the bulk of type Ia SNe. A similar
result is provided by Fig.~\ref{fig:2} for the run of [Fe/Mg] with [Fe/H]. We
remark that the values of [O/H] and [Mg/H] at which the changes of slope occur
is large ([O/H]$\sim$[Mg/H]$\sim -0.3$). 

A direct comparison with existing galactic evolution model is possible for O,
for which the predictions of MF models are available. We overposed on
Fig.~\ref{fig:1}a lines representing the predictions given by models 1 and 2 of
MF; these models were computed with an $e-$folding time of 1 Gyr for the infall
(halo collapse), and two different laws of star formation: the times required
for [Fe/H] to raise at [Fe/H]=$-1$\ (roughly corresponding to the halo-disk
transitions) are $1.5\, 10^9$\ and $3\,10^8$~yr respectively. Both models were
arbitrarily scaled to match the [Fe/O] ratio for halo stars ([Fe/H]$<-1$).
These scalings only imply small changes in the adopted yields of Fe from type
II SNe, which on turn depend on the cut-off mass for the remnants for
core-collapse SNe, an ill-defined quantity at present (Timmes et al. 1995).
While these scalings do not affect our conclusions, they help to see the main
features we like to point out. 

Insofar metal-poor stars are considered ([Fe/H]$<-0.5$), the r.m.s. values of
the residuals of points for individual stars around the lines representing the
MF models are 0.158 and 0.127 dex for model 1 and 2 respectively: data for
metal-poor stars are then better represented by model 2, which considers the
production of Fe from massive stars alone during this phase (the shallow slope
of [Fe/O] with [O/H] in this model is due to the dependence of the O/Fe
abundance ratio in the ejecta of type II SNe with progenitor mass); while model
1 is in clear disagreement with observations. The conclusion of this (and other
comparisons not shown in Fig.~\ref{fig:1}), is that the raise of O abundances
up to [O/H]=$-0.3$\ occurred on a timescale which is not much longer than the
lifetime of type Ia SNe. This timescale will be better quantified later
\footnote{While this conclusion is based on our analysis, it is not obvious
that a different conclusion would be obtained even adopting the O abundances
from the UV OH band by Israelian et al. (1998) and Boesgaard et al. 1999). In
fact, the quite large slope found by these analyses might be explained by a
very fast raise in the metal content of the early gas of the galaxy - so fast
that only the most massive stars were able to pollute the early interstellar 
medium; or by models which consider independent chemical evolution of
individual halo fragments, like that proposed by Tsujimoto et al. (1999)}.

While model 2 is able to better reproduce observations at low metallicities, it
fails in the metal-rich range. In fact, $\chi^2$-tests show that a linear
dependence of [Fe/O] on [O/H] (roughly similar to that given by both model 1
and 2) is not a good representation of the observed [Fe/O]'s for [O/H]$>-0.5$,
since the scatter of [Fe/O] values for [O/H]$<-0.1$\ is much larger than
observational errors. A very similar result might be obtained for Mg. The
simplest interpretation is that the Fe content suddenly increased at
[O/H]$\sim$[Mg/H]$\sim -0.3$; the implication is that there was a phase in
which the production of O and Mg (i.e. the formation of type II SNe) felt down
to small values, leaving only the Fe producers in activity. This obviously
means a sudden decrease in the formation of massive stars and, since we are
moving in the framework of a constant initial mass function (MF), in the star
formation {\it tout court}. We found this same feature when considering Si and
Ca abundances rather than O ones (these abundances are not discussed in Paper
III, but they display trends similar to those found for O and Mg: Gratton et
al., in preparation). The synthesis of these elements is also likely related to
massive stars (Thielemann et al. 1990).

The phase of low star formation must have lasted enough to allow the explosion
of a large fraction of halo and thick disk type Ia SNe, since [O/H] and [Mg/H]
start increasing again from values of [Fe/O] and [Fe/Mg] $\sim 0.2$~dex higher
than that achieved in the previous phase. The simultaneous increase of both
[O/H] and [Fe/O] in stars with [O/H]$>-0.5$, [O/Fe]$>-0.25$\ seems to require
that both type Ia and type II SNe contribute to the chemical enrichment during
this phase. The regression line through data in this region is: 
\begin{equation}
{\rm [Fe/O]}=(0.34\pm 0.07){\rm [O/H]} + (0.04\pm 0.10),
\end{equation}
based on 71 stars. However, results for Mg in the metal-rich regime are
quite different; the regression line through data in this region is: 
\begin{equation}
{\rm [Fe/Mg]}=(-0.055\pm 0.025){\rm [Mg/H]} - (0.089\pm 0.076),
\end{equation}
based on 164 stars. If real, this result indicates that Mg abundances do not
increase at the same rate as O ones, suggesting that the O/Mg ratio in the
ejecta of type II SNe is a function metal abundance; this could be understood
if severe mass loss reduce production of Mg in massive, metal rich stars. 

\begin{table*}
\caption[ ]{Average chemical and dynamical parameters for stellar populations
in the solar neighbourhood determined from observed stars; for each parameter
we give the number of stars used to compute the average values, the values, and
the r.m.s. scatter of individual points around the mean }
\label{tab:1}
\begin{flushleft}
\begin{tabular}{lrclrclrcl}
\hline\noalign{\smallskip}
Parameter&\multicolumn{3}{c}{Halo}&\multicolumn{3}{c}{Thick disk}&
\multicolumn{3}{c}{Thin disk}\\
\noalign{\smallskip}
\hline\noalign{\smallskip}
$V_{\rm rot}$~(\kms) &21&  ~~48 & ~100& 21 &  144  &  52 & 164 &   206 &  24 \\
$z_{\rm max}$~(Kpc)  &21&  ~2.1 & 2.3 & 21 & ~0.6  & 1.1 & 164 &  ~0.18& 0.20 \\
$e$                  &21& ~~0.69& 0.28& 21 &~~0.38 & 0.19& 164 & ~~0.14& 0.07\\
$\log t$~(Gyr)       &13& ~~1.16& 0.08& 21 &~~1.12 & 0.06& 164 & ~~0.69& 0.23\\
$[$Fe/H$]$           &29&$-$1.68& 0.41& 21 &$-$0.63& 0.15& 164 &$-$0.19& 0.26\\
$[$O/H$]$            &29&$-$1.23& 0.40& 19 &$-$0.29& 0.12& ~68 &$-$0.18& 0.18\\
$[$Fe/O$]^a$         &28&$-$0.46& 0.10& 19 &$-$0.36& 0.06& ~67 &$-$0.01& 0.11\\
$[$Mg/H$]$           &22&$-$1.42& 0.41& 21 &$-$0.29& 0.13& 164 &$-$0.10& 0.24\\
$[$Fe/Mg$]$          &22&$-$0.30& 0.10& 21 &$-$0.32& 0.05& 164 &$-$0.08& 0.08\\
\noalign{\smallskip}
\hline
\end{tabular}
\end{flushleft}
\end{table*}

\begin{figure}
\psfig{figure=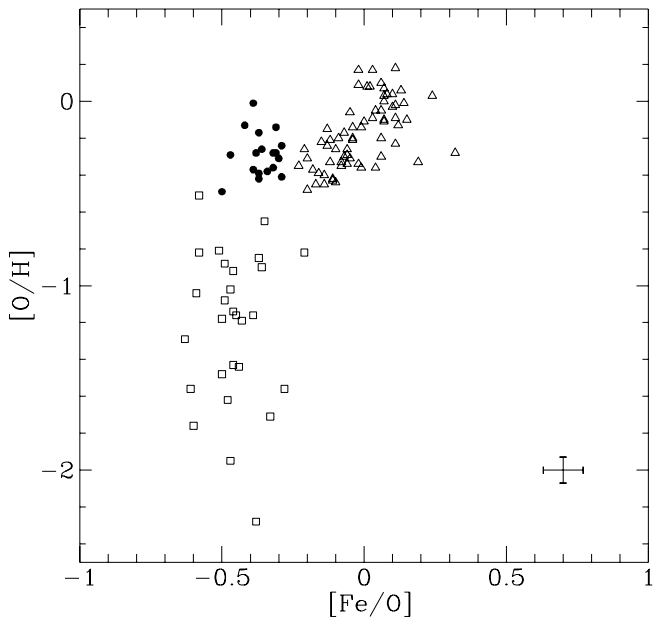,width=8.8cm,clip=}
\caption[ ]{ Run of [Fe/O] ratio with [O/H] for the stars of Figure~1
having accurate dynamical parameters. Different symbols
mark stars in different areas of the diagram. Compare the distribution of
points in this diagram with those of the next figure. Error bar is at bottom 
right} 
\label{fig:3}
\end{figure}

\begin{figure}
\psfig{figure=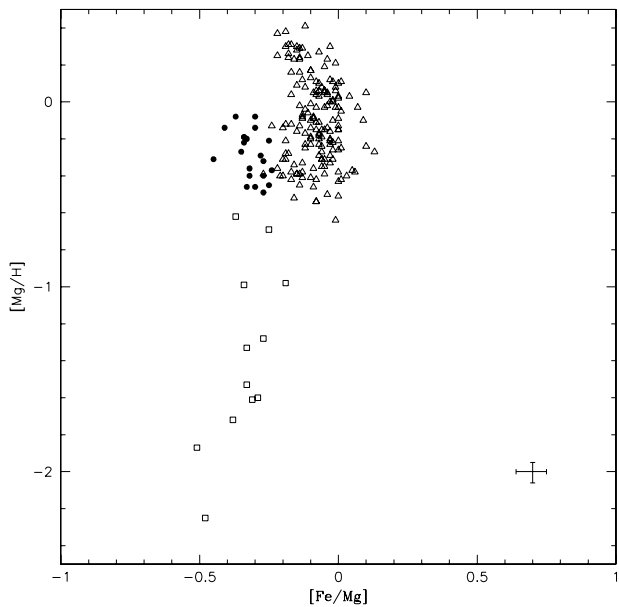,width=8.8cm,clip=}
\caption[ ]{ The same as Figure 3, but for Mg rather than for O }
\label{fig:4}
\end{figure}

\begin{figure}
\psfig{figure=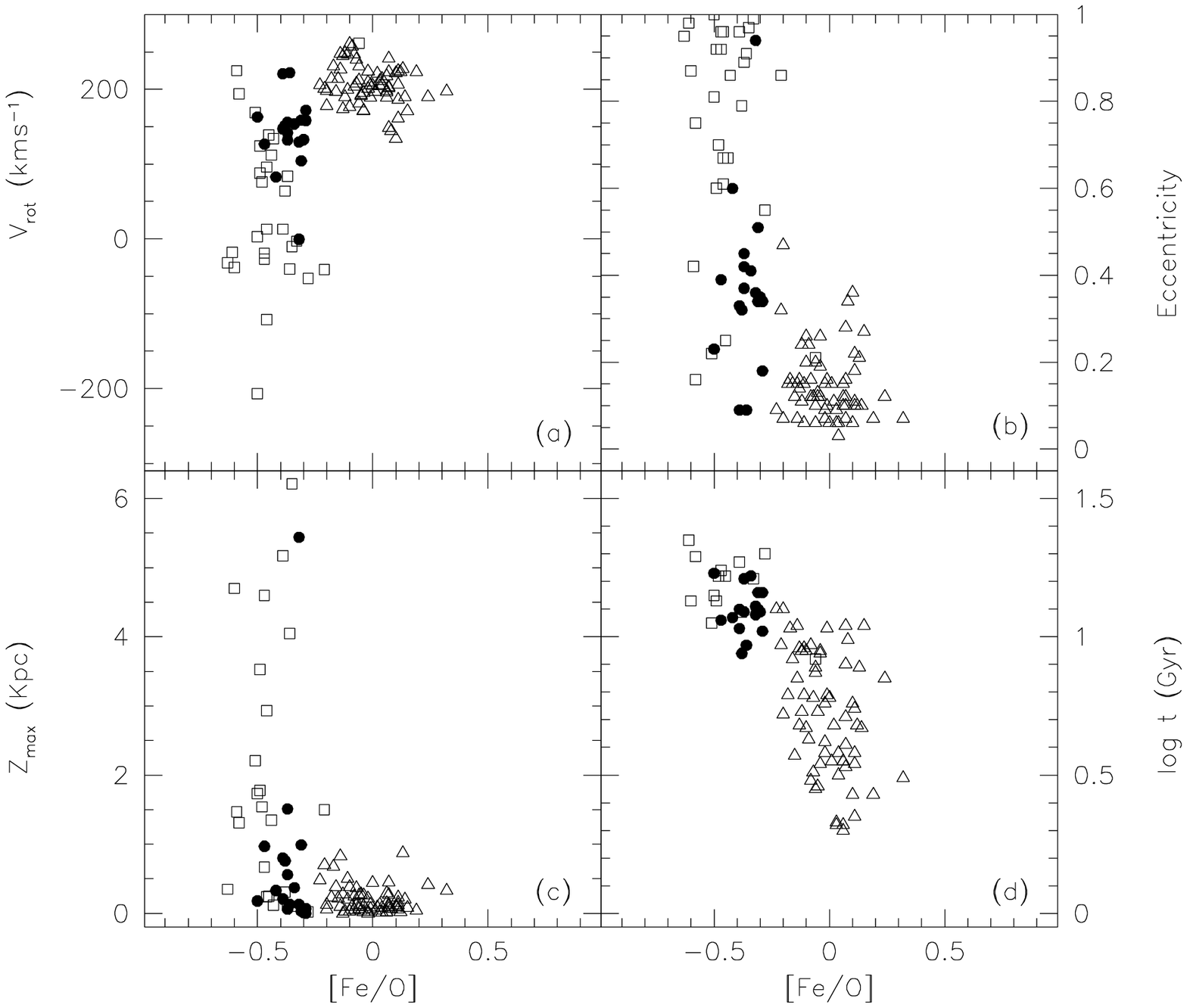,width=8.8cm,clip=}
\caption[ ]{ Comparison between the [Fe/O] ratios and physical and dynamical
parameters (rotational velocity around the galactic centre $V_{\rm rot}$: panel
a; orbital eccentricity $e$: panel b; maximum height of orbit above the
galactic plane $z_{\rm max}$: panel c; age $t$: panel d) for the stars of
Figure~4. Different symbols marks stars with different chemical abundances:
open squares are stars with [O/H]$<-0.5$; filled circles are stars with
[O/H]$>-0.5$\ and [O/Fe]$<-0.25$; open triangles are stars with [O/H]$>-0.5$\
and [O/Fe]$<-0.25$. Note the close correspondance between classification of
stellar populations in the solar neighbourhood according to chemical and
physical criteria} 
\label{fig:5}
\end{figure}

\begin{figure}
\psfig{figure=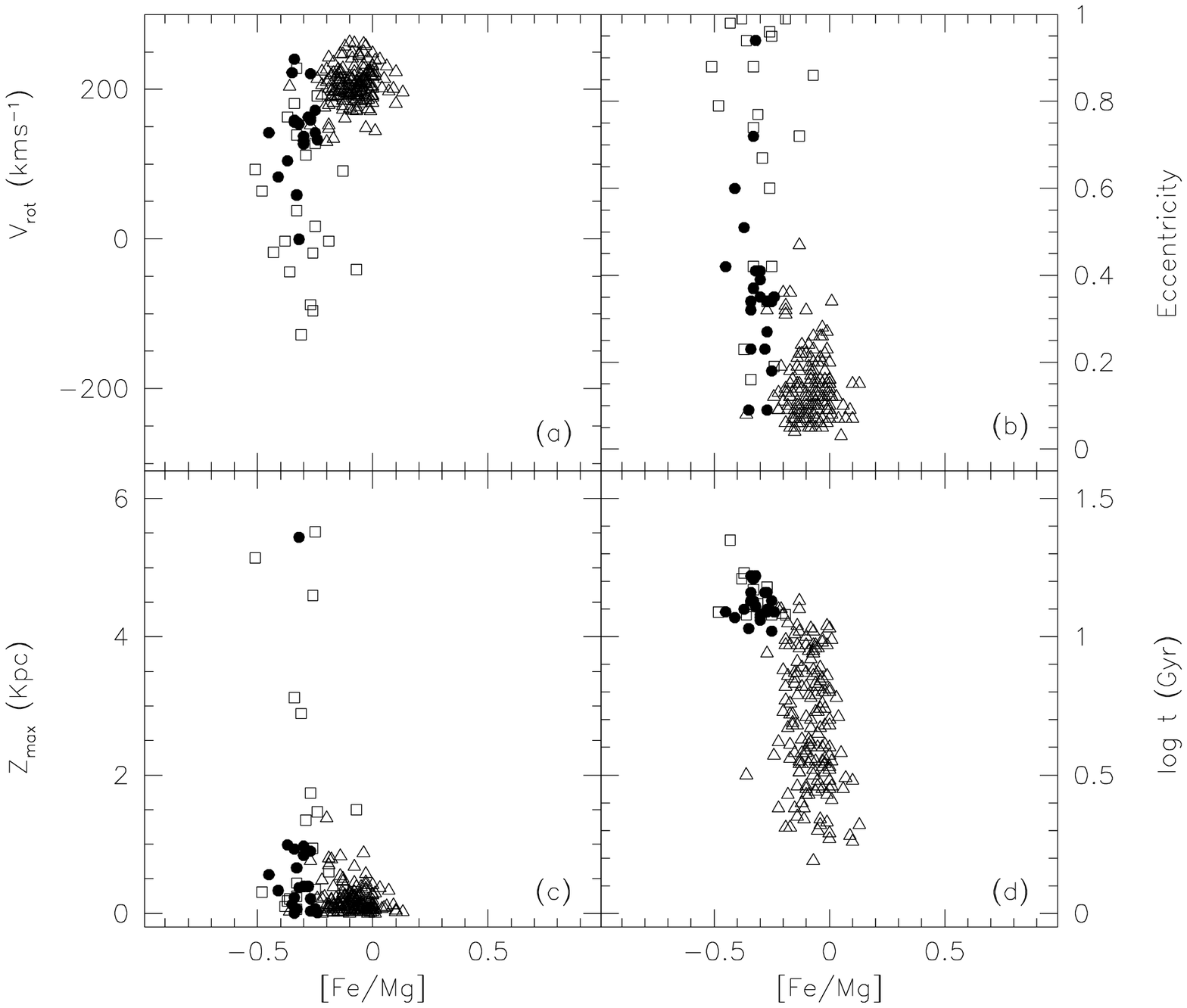,width=8.8cm,clip=}
\caption[ ]{ Comparison between the [Fe/Mg] ratios and physical and dynamical
parameters (rotational velocity around the galactic centre $V_{\rm rot}$: panel
a; orbital eccentricity $e$: panel b; maximum height of orbit above the
galactic plane $z_{\rm max}$: panel c; age $t$: panel d) for the stars of
Figure~4. Different symbols marks stars with different chemical abundances:
open squares are stars with [Mg/H]$<-0.6$; filled circles are stars with
$-0.5<$[Mg/H]$<0$\ and [Mg/Fe]$<-0.25$; open triangles are the remaining metal
rich stars. Note the close correspondance between classification of stellar
populations in the solar neighbourhood according to chemical and physical
criteria } 
\label{fig:6}
\end{figure}

\begin{figure}
\psfig{figure=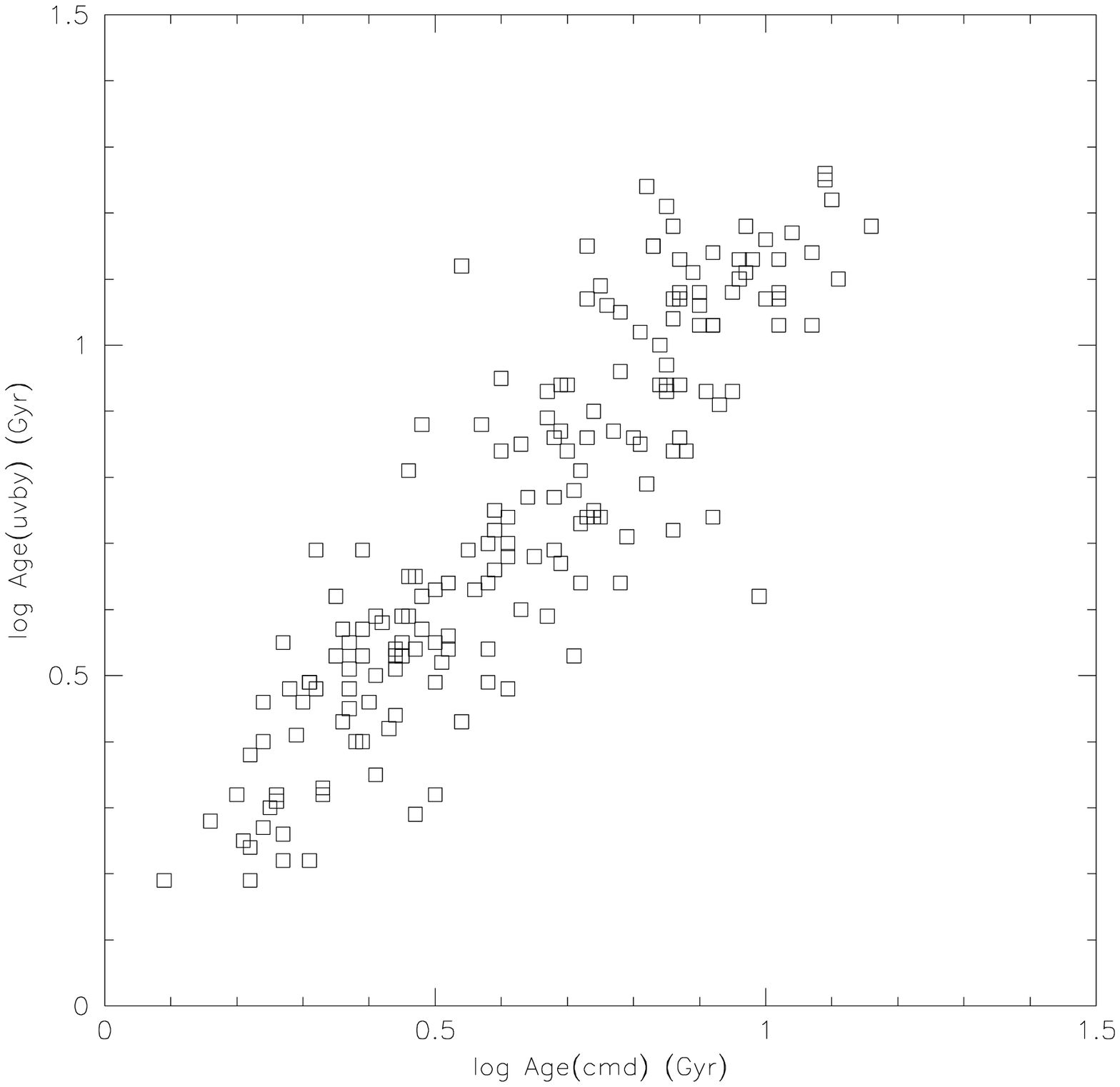,width=8.8cm,clip=}
\caption[ ]{ Comparison between the ages listed by Edvardsson et al. (1989),
and the values determined from the location in the colour-magnitude diagram
using absolute magnitudes and colours from HIPPARCOS (Perryman et al. 1997) } 
\label{fig:10}
\end{figure}

\section{ Chemical abundances and dynamics }

A further basic step can be done by combining information provided by the
([O/H] vs [Fe/O]) and ([Mg/H] vs [Fe/Mg]) diagrams with those obtained from the
kinematics of stars in our sample. A caveat should be done here, since stars
analyzed in the present paper were collected from various sources, and the
(sometimes not well defined) selection criteria introduce important biases: the
distribution of stars with O and Mg abundances in our sample is very different
from that obtained from a volume limited sample in the solar neighbourhood, and
high velocity stars are likely overrepresented amongst the most metal poor
ones. In this section we will use this comparison simply to identify stellar
populations defined on chemical grounds with those defined from dynamics. 

The basic data for this comparison are the O, Mg and Fe abundances drawn from
our analysis, and dynamical data determined by Edvardsson et al. (1993), for
stars with [O/H]$>-0.5$; for stars more metal-poor than this limit in the
Tomkin et al. (1992) and Zhao \& Magain (1990) samples, similar data were
obtained using parallaxes and proper motion from HIPPARCOS (Perryman et al.
1997), radial velocities from the literature (Carney et al. 1994 whenever
possible; else data were taken from the SIMBAD database), and the code for
galactic orbit calculations by Aarseth as modified by Carraro (1994). Both
these samples only include dwarfs in the solar neighbourhood; the Edvardsson
et al. sample is essentially a magnitude-limited sample (although the
magnitude limit is a - known - function of metallicity), with no kinematical
bias; while Tomkin et al. (1992) and Zhao \& Magain (1990) selected high
proper motion stars, so that a bias toward high velocity stars is certainly
present amongst the most metal-poor stars.

The location of the stars with known dynamical data in the ([O/H] vs [Fe/O])
diagram is shown in Fig.~\ref{fig:3}, where we plotted with different
symbols stars in different regions of the diagram: metal-poor stars
([O/H]$<-0.5$: group A); Fe-poor O-rich stars ([O/H]$>-0.5$, [Fe/O]$<-0.25$:
group B); and Fe-rich O-rich stars ([O/H]$>-0.5$, [Fe/O]$>-0.25$: group C). The
analogous diagram for Mg is shown in Fig.~\ref{fig:4}, but in this case
group A are stars with [Mg/H]$<-0.6$; group B are stars with $-0.5<$[Mg/H]$<0$\
and [Mg/Fe]$<-0.25$; and group C are the remaining metal rich stars. A
star-by-star comparison shows that stars are attributed to the same groups
when using O and Mg.

In the four panels of Fig.~\ref{fig:5} and ~\ref{fig:6} [Fe/O] and [Fe/Mg]
ratios for these stars are plotted against the rotational velocity around the
galactic center, the orbital eccentricity, the maximum height of the orbit
over the galactic plane $z_{max}$, and the age $t$\ respectively. The age
values are averages of two independent values:
\begin{itemize}
\item those determined by Edvardsson et al. (1993) and Schuster \& Nissen 
(1989) derived using an homogenous procedure from the $(b-y)-c_1$ diagram, 
calibrated against age using standard isochrones ({\it i.e.} not O-enhanced) 
by VandenBerg and Bell (1985)
\item and those derived using absolute magnitudes from HIPPARCOS parallaxes
(Perryman et al. 1997), interpolating within the grid of isochrones
by Padua group (Girardi et al. 2000); in this case the enhancement of O and
the other $\alpha-$elements was considered by modifying the Fe abundance
according the procedure suggested by Straniero et al. (1997)
\end{itemize}
The two age scales are somewhat different; they were homogeneized to an a
common (arbitrary) scale before averaging them.

Absolute ages are affected by various uncertainties related to stellar models,
while the relative ranking should be quite reliable, as indicated by the
good agreement existing between the two set of age determinations (see 
Fig..~\ref{fig:10}). Since basic data for the
two age estimates are independent each other, internal uncertainties on the
ages may be obtained by the r.m.s scatter of the differences; in this way, we
estimate that typical error bars in log(Age) for individual stars are $\pm
0.06$~dex (i.e. $\pm 15$\%). The internal scatter we get for group A and B is
of this same order, indicating that small if any internal scatter exists for
these groups.

The average properties for the three chemical groups are listed in
Table~\ref{tab:1}. 

The main conclusions we may draw from Fig.~1-6 and Table~\ref{tab:1} are:
\begin{enumerate}
\item Though some bias may be present, present data support the identification
of group A as the halo, of group B as the thick disk (see e.g. Robin et al.
1996, Norris, 1999, Buser et al. 1999), and of group C as the thin disk.
\item In the framework of homogeneous models for the galactic evolution, the
formation of stars in the halo and the thick disk was fast (a few $10^8$~yr),
i.e. shorter than the typical timescale  of evolution for the progenitors of
type Ia SNe. 
\item Given the lack of a clear selection criterion amongst metal-poor stars,
our data cannot be used to draw any conclusion about clear breaks
between the halo and thick disk populations; the age difference must be small
(i.e. the two populations are virtually coeval), else the [O/Fe] and [Mg/Fe]
ratios of the two populations should be different; the ages derived for both
groups are well in excess of 10~Gyr. Anyway, this discontinuity is supported by
other studies (see Norris 1993). As noticed by Wyse and Gilmore (1992), the
dynamical properties of the thick disk are clearly indicative of a flattened
intermediate population supported by rotation (these characteristics are drawn
from the Edvardsson et al. sample, which are not affected by kinematical
biases); while the halo space distribution is much wider, and the system
appears to be supported by velocity dispersion rather than by rotation.
In our analysis, this last result is likely enhanced by the selection biases
in the Carney et al. sample. Further strong supports to the present conclusion
that the thick disk is a homogeneous, chemically old population is given by the
results by Fuhrmann (1998, 1999), and Nissen \& Schuster (1997): these authors
found that thick disk stars have low [Fe/O] and [Fe/Mg] ratios (equal or even
lower than those found for halo stars of similar metallicity), with very small
intrinsic scatter.
\item There was a sudden decrease in star formation during the transition
between the thick and thin disk phases, which are then clearly distinct at
least from the chemical point of view. The phase of low star formation must
have lasted at least 1~Gyr in order to allow for the explosion of the bulk of
type Ia SNe; however, the hiatus was not longer than 3~Gyr, else there would
be obvious break in the [Fe/O] and [Fe/Mg] vs age diagrams. We remark that the
age difference between the thick disk and the oldest thin disk stars could have
been overestimated in Fig.~\ref{fig:5}d and ~\ref{fig:6}d, since these ages
were derived assuming solar ratios of O and Mg to Fe. Again, results from 
Fuhrmann (1998, 1999) and Nissen \& Schuster (1997) well support this picture.
\item Both the [Fe/O] and [Fe/Mg] ratios for thin disk stars increase with
time: in fact there is a significative correlation of these ratios with stellar
ages; the linear regression line are: 
\begin{equation}
{\rm [Fe/O]} = -(0.19\pm 0.06) \log t + (0.12\pm 0.11),
\end{equation}
based on 68 stars for O, and: 
\begin{equation}
{\rm [Fe/Mg]} = -(0.085\pm 0.026) \log t - (0.024\pm 0.075),
\end{equation}
based on 164 stars for Mg. The Persson correlation coefficients are 0.35 and
0.25 respectively: the probability of getting such high correlation
coefficients by chance are $<<0.005$ and $<<0.01$. The scatter around the mean
lines are compatible with the error bars in ages, [Fe/O] and [Fe/Mg] values. 
\end{enumerate}

\begin{figure}
\psfig{figure=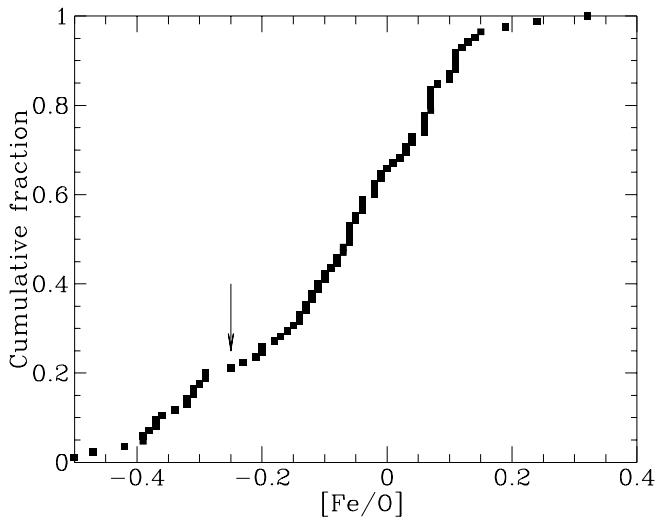,width=8.8cm,clip=}
\caption[ ]{ Cumulative distribution function of [O/Fe] ratios amongst stars
in the Edvardsson et al. sample. Note the possible presence of a gap at
[Fe/O]$\sim -0.25$} 
\label{fig:7}
\end{figure}

\begin{figure}
\psfig{figure=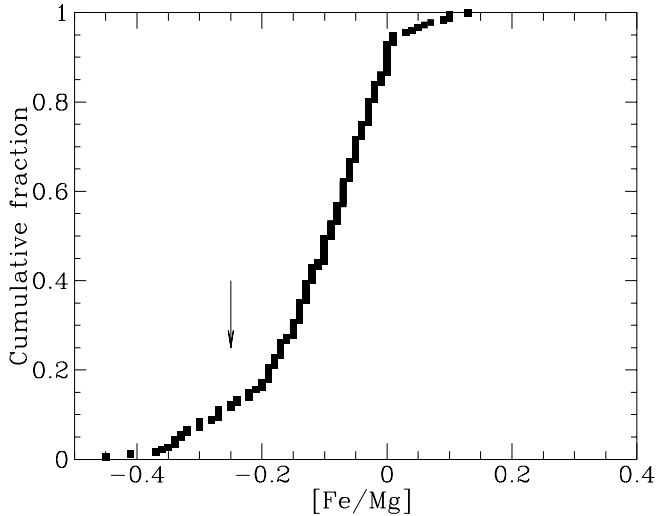,width=8.8cm,clip=}
\caption[ ]{ The same as Figure~7, but for Mg rather than for O. Note again the
possible presence of a gap at [Fe/Mg]$\sim -0.25$ }
\label{fig:8}
\end{figure}

\section{ The gap in [Fe/O] distribution }

If the raise of [Fe/O] at constant [O/H] is due to a sudden decrease in star
formation during the transition between the thick and thin disk phases, then
there should be a corresponding gap at [Fe/O]$\sim -0.25$ in the distribution
of stars with [Fe/O]. A cumulative diagram of the distribution of [Fe/O] and
[Fe/Mg] values for stars in the Edvardsson et al. sample (Fig.~\ref{fig:7} and
~\ref{fig:8})
indeed supports this inference. In the rest of this section, we will evaluate
the statistical significance of this gap for O; similar (though slightly less
significative) results were obtained for Mg. To this purpose, we performed
tests on the distribution of [Fe/O] values in that sample, which consists of
the brightest {\it bona fide} single stars with $5600<$\teff$<6800$~K, having
an absolute magnitude from 0.4 to 2~mag smaller than that of the Zero Age Main
Sequence at the same $b-y$\ colour, in equally spaced bins in [Fe/H].
Generally, populations of each bin were kept similar, but the two most
metal-poor bins ([Fe/H]$<-0.6$) have populations about half those of the
others, due to the magnitude limit of the survey. However, the 85 stars for
which O abundances are available are nearly uniformly distributed with [Fe/H],
since we have these data for 42\% of the stars with [Fe/H]$>-0.6$, and for 79\%
of the stars more metal-poor than this limit. A $\chi^2$-test confirms that the
distribution of stars having [Fe/O] values cannot be distinguished from a
uniform distribution in [Fe/H]; anyway, we repeated the following analysis
using both a uniform distribution, and distributions obtained by summing a
random gaussian distributed term (representing observational errors) to the
observed [Fe/H]'s. The results are very similar. 

We first tested the hypothesis that the [Fe/O] values are distributed
uniformly, as expected if a linear dependence of [Fe/O] on [Fe/H] holds. We
found that this hypothesis can be rejected at a high level of confidence, using
both a $\chi^2$-test and a serie of Monte Carlo simulations. Each Monte Carlo
simulation included 10,000 extractions of 85 [Fe/H] values (using both early
described approaches); individual [Fe/O]'s were the sum of the values deduced
from these [Fe/H]'s using the [Fe/O]-[Fe/H] law, and of a random gaussian
distributed term representing the spread due to observational errors and to the
intrinsic star-to-star variations. The simulations were repeated for three
values of the standard deviation for this term: 0.05, 0.07, and 0.10~dex; the
intermediate value was deduced from our error analysis, and it is equal to the
standard deviation from the mean value for thick disk stars; the last one is
consistent with the residuals around the best fit relation for thin disk stars.
Most of the deviation from a uniform distribution in [Fe/O] is due to the
excess of stars with [Fe/O]$<-0.3$\ (corresponding to the thick disk
population), and to the lack of stars with $-0.3<$[Fe/O]$<-0.15$\ (the expected
location of the gap). It can be noticed that a quadratic dependence of [Fe/O]
on [Fe/H] gives a better fit to observational data. We then repeated the Monte
Carlo simulations with a similar [Fe/O]-[Fe/H] law. We found that the
probability that the very low number of stars in a bin of 0.1~dex centered at
[Fe/O]=$-0.22$\ is due to chance is about 0.024 for a uniform distribution with
[Fe/H], and 0.013 for a distribution function equal to the frequency
distribution (the exact values depend on the assumed observational errors; the
above mentioned values refer to the less significant cases obtained assuming
that present [Fe/O] values have errors of 0.10~dex). A good significance
(chance probability $<0.05$) is achieved for 0.1 dex bins centered over the
range $-0.22<$[O/Fe]$<-0.25$; this suggests that the gap is broader than
0.1~dex. 

We conclude that present available data support the hypothesis that there is a
gap in the distribution of [Fe/O] as expected from a sudden decrease in star
formation during the transition from the thick to thin disk phases; however we
think this test should be repeated using a larger and properly selected sample.

\section{ Comparison with models of galactic chemical evolution }

The overall run of [Fe/O]\footnote{In this section we will only use Fe and O
abundances; however, results obtained by considering Mg abundances would be
very similar.} with [Fe/H] is certainly related to the delayed Fe synthesis:
however, the interplay between star formation rate, progenitor lifetime, and
infall can only be cleared out by detailed modelling of galactic chemical
evolution. Unfortunately, various aspects of star formation and evolution are
still not well understood, so that these models have several free parameters.
Furthermore, a fully appropriate comparison between abundances in metal-poor
stars and models of galactic evolution should require models which include
consistent and detailed treatment of both chemistry and dynamics. However, this
is still beyond current computational capabilities, and at present dynamics has
to be introduced into chemical evolution models parametrically, increasing even
more the number of free perameters. In order to explore allowed ranges for some
of the relevant quantities, we then computed a large number of single-zone
models (with infall of original unprocessed material) appropriate for the solar
neighborhood; the code we used is an improved version of that of MF. The main
advantage of this approach is the reduction in the number of free-parameters,
although results of our comparisons strictly apply only to homogeneous collapse
models, and not e.g. to accretion scenarios. We will see in the next section
that even with these limitations, results of our comparisons are enough to
suggest that the best scenario required to explain the star formation history
in the solar neighborhood should include both dissipational collapse and
accretion. 

When comparing the predictions of our models with observations, we considered a
wide range of constraints, including present abundances in the interstellar
medium, current gas density, star formation and SN rates, the distribution of
abundances among G-dwarfs, the run of [Fe/H] with age, IMF, etc (Matteucci
1991), and retained only those models which give an overall match to all of
them. We find that in the best cases, r.m.s. values of the residuals of
[Fe/O]'s for individual stars around lines representing our models are $\sim
0.10$~dex for halo and thick disk stars, and $\sim 0.08$~dex for thin disk
stars, in good agreement with observational errors, as shown by Monte Carlo
simulations where both errors in [Fe/O] and [O/H] were taken into account
(similar simulations were also used to reject those models providing poor
fits). Predictions about the [Fe/O] ratios provided by some of these models are
plotted in Fig. ~\ref{fig:3}b and c, overposed to the same observational points
shown in Fig.~\ref{fig:3}a. 

On the whole, we found that only models having two distinct infall episodes,
the first connected to the halo (and thick disk), and the second to the thin
disk, fit observational data (see also Chiappini et al. 1997 for similar
models); while models with a single infall episode do not give a good match
even when star formation law was changed with time, due to either large
accumulation of gas during the phase corresponding to the raise of the [Fe/O]
ratio, or lack of gas at present (depending on the rate of decay). Decay of
the halo infall rate should be fast in order to reproduce the almost flat run
of [Fe/O] with [O/H] in the halo and thick disk: the $e-$folding time is
shorter than or equal to the adopted time delay for type Ia SNe, i.e. $\sim
0.5$~Gyr. The halo infall should have contributed no more than 20-30\% of the
present density in the solar neighborhood (and the fraction in stars should be
about half that). The decay of the thin disk infall should be rather slow
($\geq 4$~Gyr, which implies a present infall rate $\geq
1~\Msol~$pc$^{-2}$Gyr$^{-1}$), else there would be not enough gas at present,
the present interstellar medium should be too metal rich, the observed roughly
linear run of [Fe/O] with [O/H] in the thin disk would not be reproduced, and
there would be an excess of moderately metal-poor stars ([Fe/H]$\sim -0.5$).
This value for the present infall rate is larger than the upper limit deduced
from observations ($\sim 0.7~\Msol~$pc$^{-2}$Gyr$^{-1}$: Mirabel 1989). This
inconsistency might be removed by either assuming that star formation law
changes with time (the efficiency should then be larger in the halo than in the
thin disk), or that the initial mass function (IMF) is flatter at large masses
(slope $\sim 1.3$) than that adopted in most of our models (Scalo 1986) (note
that in order to avoid O overproduction with this flatter IMF, an upper limit
of $40\div 50~\Msol$\ has to be adopted for type~II SNe). Both these
alternatives do not contradict basic costraints.

Models where both infall episodes start at high values at the beginning and
then decay (Fig.~\ref{fig:3}b), as well as models where the disk infall is
delayed with respect to the halo one (Fig.~\ref{fig:3}c), fits data quite well.
It should be noticed that in the first class of models the star formation must
be arbitrarily lowered at the thick-thin disk transition (that should occur
$\sim 1$~Gyr after beginning), and there is no hiatus: the raise of the [Fe/O]
ratio at constant [O/H] is due to Fe production by the large number of type Ia
SNe from the halo-thick disk, while in the meantime only small amounts of O are
produced, due to the decreased star formation rate, which barely compensate for
the dilution by metal-poor infalling material (the [O/H] ratio starts
increasing again when most gas is consumed and the infall rate decays down to
small values). The less appealing aspect of this class of models is the large
fraction of metal-poor stars, that only can be reconciled with the small number
of metal-poor G-dwarfs observed in the solar neighborhood by assuming that the
scale height is a strong function of metallicity. On the other side, the second
class of models naturally yield to a hiatus if a threshold (surface) gas
density for star formation is adopted (Kennicutt 1989). A threshold gas
density might be expected if the processes involved in star formation are
self-regulating, and for a low enough surface gas density the feedback
mechanisms that regulate the star formation rate break down (Gallagher \&
Hunter 1984; Elmegreen 1992; Burkert et al. 1992). Observationally, an
approximate value of $1\div 2\ 10^{21}$~atoms cm$^{-2}=5\div
10~\Msol~$pc$^{-2}$\ has been proposed for this threshold from observation of
irregular and HI-rich spiral galaxies (Skillman 1986; van der Hulst et al.
1987); this threshold value is close to the current gas density in the solar
neighbourhood (Rana \& Basu 1991; Kujiken \& Gilmore 1989), as expected in a
self-regulating mechanism. With the above mentioned infall rates, a star
formation rate with threshold produces a hiatus in star formation at the end
of the halo and thick disk phase; in our models, star formation in this early
phase lasts for 1.5-2.5 Gyr (inversely depending on the adopted threshold
density), and for given yields the final [O/H] value depends on the ratio
between the halo infall and the threshold density, because the early chemical
evolution is essentially that of a closed box (Phillips et al 1990). In these
models, type Ia SNe begin to contribute to nucleosynthesis during the last
phases of halo and thick disk evolution, raising the [Fe/O] ratio; this
contribution continues during the hiatus, which must be at least as long as
time delay of SN Ia ($\geq 0.5$~Gyr), in order to raise the [Fe/O] by at least
0.13 dex, that we estimate as the lower limit given by observations. However,
the upper limit for hiatus duration cannot be determined from the [Fe/O] run
due to a saturation effect.

Part of the gas might be lost at the thick-thin disk transition (due e.g. to a
galactic wind induced by SN explosions: Silk 1985); if this occurs, this
metal-rich gas will be replaced by more metal-poor infalling gas, and the
starting value of [O/H] in the thin disk is lower than the maximum achieved
during the thick disk evolution. Models where up to 75\% of the gas is lost at
this epoch fit data quite well, but they predict the existence of some very
O-poor, Fe-rich thin disk stars, which are not present in the observed samples
\footnote{ A few stars with such a chemical composition are indeed known, but 
they have extreme halo kinematics (King 1997; Carney et al. 1997)}.
We think that the overposition of the thick and thin disk sequences over some
range in [O/H] can be better explained by noting that stars currently in the
solar neighborhood likely formed over a range of galactocentric distances (see
Fran\c cois and Matteucci 1993), where the halo surface density and the largest
[O/H] values achieved during the halo and thick disk evolution were different. 
Note that this is also the explanation favoured by other authors (e.g. 
Edvardsson et al. 1993).

\section{ Dissipational collapse and accretion }

What our data tells us about scenarios of Galaxy formation? The fast rate of
star formation for the halo and thick disk ($<10^9$~yr) is consistent with a
smooth dissipational collapse (Eggen et al. 1962; Larson 1974). However, in
this case the transition from the thick to thin disk phases should be
continuous: some heating mechanisms causing the observed sudden decrease in
star formation during this transition is missing and should then be
introduced. Possible candidates are a high SN rate (Silk 1985; Berman \&
Suchkov 1991), and merging of smaller galaxies with our own; these mechanisms
might also be acting simultaneously.

Our tests with models with a single infall episode (but variable star formation
rate) suggest that a simple heating as that produced by SNe cannot reproduce
the whole spectrum of observations. Also, the discontinuity in specific 
angular momentum between halo and thick disk suggests that there is not a
smooth transitions between these two populations (Wyse and Gilmore 1992).
On the other side, merging with gas poor
satellite(s) having a mass larger than a few hundredths the disk mass may heat
(and destroys) any pre-existing stellar thin disk (Quinn et al. 1993, Walker
et al. 1996) and create a thick disk, although this result is not obtained in
all simulations (see e.g. Huang \& Carlberg 1997), so that it seems to depend
on the initial conditions as well as on the properties of the satellite:
merging of satellites on prograde orbits more likely produce thick disks, while
those on retrograde orbits mainly produce disk tilts (Velazquez and White
1999). The mass range is fixed by the amount of kinetic energy to be injected
into the disk: merging with a companion of comparable mass would have
transformed the Milky Way into an elliptical, while merging with a very small
satellite has only minor effects. If the disk or the satellite contained gas
(as indicated by chemical evolution models), merging was likely accompanied by
a burst in star formation; a possible support in favour of such a burst is the
presence of a numerous population of globular clusters likely connected to the
thick disk, distinct from the population connected to the halo (Zinn, 1985).
Perhaps this burst exhausted the existing gas, contributing to the present
thick disk or it may have been concentrated in the bulge regions, far from the
solar neighbourhood (as suggested by some simulations: Mihos \& Hernquist
1994), or finally the high SN rate might have caused a galactic wind (Berman
\& Suchkov 1991). Anyway, the present thin disk should have formed later by a
secondary process (Ostriker \& Thuan 1975); models of disk evolution
(Kennicutt 1989; Burkert et al. 1992) then indicate that some time would be
required before the critical density was reached and star formation started
again. Thick disk stars are then likely older than the oldest stars with
thin-disk kinematics and a discontinuity would be expected between the thick
and thin disk phases. This agrees with our Fig.~\ref{fig:5}d, which also
indicates that a similar large merging could not have occurred during the last
10 Gyrs, although observations of the Sagittarius dwarf galaxy (Ibata et al.
1994), presently merging with the Milky Way, indicate that minor episodes are
still occurring. The n-body simulations (Quinn et al. 1993) indicate that if
relatively large merging occurred, the present thick disk would be composed of
both stars belonging to the merged satellite(s) and to the original disk; the
last one should dominate due to its larger mass.

A pure accretion scenario readily explains the lack of appreciable kinematic
and chemical gradients in the outer halo (Carney 1993), but it fails to explain
the gradients observed in the inner regions. In a smooth dissipational collapse
scenario, like that of Eggen et al., the proposed merging episode appears as an
{\it ad hoc} hypothesis. However, more realistic simulations of galaxy
formation by dissipational collapse which include dark matter, gas dynamics,
star formation and SN feedback (Katz 1992) suggest that many stars form in the
cores of dark matter clumps that form during the collapse: even within this
scheme then a long-living thin disk could likely form only after the end of an
early chaotic phase. The emerging favoured scenario considers then the
inhomogeneous dissipational collapse of the protogalaxy with formation of a few
secondary fragments (having of the order of several hundredths or even a few
tenths of the total mass) which are accreted later, as proposed by Norris
(1994). In this scenario, a significant fraction of the halo is due to
accretion of fragments, explaining its low specific angular momentum (Wyse and
Gilmore 1992). Our contributions to this scheme is the consideration that the
similar [Fe/O] ratios for thick disk and halo stars can best be understood if
timescales for both contraction and accretion are $\leq 1$~Gyr (although some
later accretion of low-mass fragments is possible), and the suggestion that
this scheme might naturally produce a discontinuity between the thick and thin
disk phases.

Up to now, the proposed scenario refers to the Milky Way. However, it can be
easily extended to other galaxies. A strong support to a scenario of spiral
formation including both dissipational collapse and accretion is given by the
observation that spirals without significant bulges do not have thick disks
(van der Kruit \& Searle 1981a, 1981b; Morrison et al. 1994, 1997; Fry et al.
1999; Matthews et al. 1999). Then (i) the presence of thick disks and bulges
is not an obvious outcome of galactic formation, but rather depend on some
mechanism (e.g. accretion) that may or may not be active; and (ii) their
origins are likely related (although likely not on an evolutionary sequence,
due to the very different specific angular momentum). This last assertion
agrees with the low [Fe/Mg] ratios for stars in the bulge of our own Galaxy
(McWilliam \& Rich 1994), which suggest that the difference between the ages
of the bulk of stars in the bulge, halo and thick disk is $\leq 1$~Gyr. The
presence in the galactic bulge of stars much more metal-rich than the thick
disk stars in the solar neighborhood might be understood by assuming either
that these stars formed from metal-enriched material in the burst induced by
the same merging episode(s) causing the formation of the thick disk (Mihos \&
Hernquist 1994); or simply by a pre-existing radial metallicity gradient in
the early disk, which is predicted by dissipational collapse models (Larson
1974) and should not be canceled by later merging episode(s) (Quinn et al.
1993).

Within the mixed scenario, significant thick disks and bulges are related to
accretion of satellites. Currently available statistics (Zaritsky et al. 1993)
indicate that there is about one satellite with $M_B<-15$\ per primary, though
we notice that the Milky Way has a much larger number of faint companions and
the statistics is based on regions where galaxy density is lower than in the
Local Group; furthermore, there is an excess of close satellites (separation
$<50$~kpc) near the minor axis of primaries (Holmberg 1969; Zaritsky et al.
1997), and it has been suggested that the present population of satellites to
the Milky Way represents only a small fraction of the original population (see
e.g. Klypin et al. 1999). The excess of close satellite near the minor axis
may be explained by assuming that close satellites on low inclination prograde
orbits have smaller chance to survive, due to dynamical friction; it has been
suggested that these {\it missing} satellites have merged into the primary 
(Zaritsky \& Gonzalez 1999); their perturbation may have created the thick 
disk, and their gas may have fueled the bulge. Given the small number of
satellites for each primary galaxy, a strong stochastic variation from galaxy
to galaxy is expected. The present scenario is then coherent to a picture
where the entire Hubble sequence from Sc to ellipticals might be reproduced by
assuming an increasing importance of accretion, which should be correlated
with the density of galaxies in the local environment (see e.g. Schweizer
2000).

\acknowledgements{ We wish to thank Dr G. Carraro and S. Aarseth for having
provided a copy of their code for calculation of galactic orbits; and
Dr S. Ryan for interesting comments on an
early version of this manuscript. } 

%
%
%
%
%
%
%

\end{document}